\documentclass[twocolumn,11pt]{aastex631}
\usepackage{graphicx,amsmath}
\usepackage{color}
\usepackage{booktabs}
\usepackage{bm}

\usepackage{times}
\usepackage{float}
\usepackage{ulem}
\usepackage{multirow}
\usepackage{amsmath}

\newcommand{\adsurl}[1]{\href{#1}{ADS}}
\providecommand{\url}[1]{\href{#1}{#1}}

\newcommand{\be}{\begin{equation}}
\newcommand{\ee}{\end{equation}}
\newcommand{\bea}{\begin{eqnarray}}
\newcommand{\eea}{\end{eqnarray}}

\usepackage{color}
\usepackage{ifthen}
\newboolean{editorial}
\setboolean{editorial}{true}
\newcommand{\editorial}[2]{\ifthenelse{\boolean{editorial}}{\textcolor{red}{[\textsf{\textbf{{#1}}}: }\textcolor{blue}{\textsf{{#2}}}\textcolor{red}{]}}{}}

\shorttitle{Origin of sBBHs via gravitational lensing}
\shortauthors{Chen, Lu \& Zhao}

\begin{document}

\title{
Constraining the origin of stellar binary black hole mergers by detections of their lensed  host galaxies and gravitational wave signals
}

\correspondingauthor{Youjun Lu}
\email{luyj@nao.cas.cn}

\author[0000-0001-7952-7945]{Zhiwei Chen}
\affiliation{National Astronomical Observatories, Chinese Academy of Sciences, 20A Datun Road, Beijing 100101, China}
\affiliation{School of Astronomy and Space Sciences, University of Chinese Academy of Sciences, 19A Yuquan Road, Beijing 100049, China}

\author[0000-0002-1310-4664]{Youjun Lu}
\affiliation{National Astronomical Observatories, Chinese Academy of Sciences, 20A Datun Road, Beijing 100101, China}
\affiliation{School of Astronomy and Space Sciences, University of Chinese Academy of Sciences, 19A Yuquan Road, Beijing 100049, China}
\author[0000-0002-6154-4381]{Yuetong Zhao}
\affiliation{National Astronomical Observatories, Chinese Academy of Sciences, 20A Datun Road, Beijing 100101, China}
\affiliation{School of Astronomy and Space Sciences, University of Chinese Academy of Sciences, 19A Yuquan Road, Beijing 100049, China}

\begin{abstract}
A significant number of stellar binary black hole (sBBH) mergers may be lensed and detected by the third generation gravitational wave (GW) detectors. Their lensed host galaxies may be detectable, which thus helps to accurately localize these sources and provide a new approach to study the origin of sBBHs. In this paper, we investigate the detectability of the lensed host galaxies for the lensed sBBH mergers. We find that the detection fraction of the host galaxies to the lensed GW events can be significantly different for a survey with a given limiting magnitude if sBBHs are produced by different mechanisms, such as the evolution of massive binary stars, the dynamical interactions in dense star clusters, and that assisted by active galactic nuclei or massive black holes. Furthermore, we illustrate that the statistical spatial distribution of those lensed sBBHs in its hosts resulting from different sBBH formation channels can be different from each other. Therefore, with the third generation GW detectors and future large scale galaxy surveys, it is possible to independently constrain the sBBH origin via the detection fraction of those lensed events with identifiable lensing host signatures and/or even constrain the contribution fractions from different sBBH formation mechanisms.
\end{abstract}

\keywords{Gravitational wave astronomy (675) --- Gravitational wave sources (677) --- Gravitational lensing (670) --- Black holes (162) --- Galaxies (573)}


\section{Introduction}
\label{sec:intro}

Detections of gravitational waves (GWs) from stellar binary black holes (sBBHs) by the Laser Interferometer GW Observatories (LIGO) and VIRGO prove the existence of a large number of sBBHs in the universe that could not be seen by electromagnetic (EM) waves \citep{2016PhRvL.116m1102A, 2019PhRvX...9c1040A, 2020arXiv201014527A, 2021arXiv211103606T,2021arXiv211103634T}. These sBBHs can be originated from (1) the evolution of massive binary stars \citep[hereafter denoted as the EMBS channel, e.g.,][]{2012ApJ...759...52D,2013ApJ...779...72D,2015ApJ...806..263D,2016MNRAS.462.3302E,  2016Natur.534..512B,
2018MNRAS.480.2011G, 2019MNRAS.486.2494G}, (2) the dynamical interactions in dense stellar systems \citep[hereafter the dynamical channel, e.g.,][]{1993Natur.364..423S, 2000ApJ...528L..17P, 2016ApJ...824L...8R}, or (3) the active galactic nuclei (AGN) or massive black hole (MBH) assisted mechanisms \citep[hereafter the AGN-MBH channel, e.g.,,][]{2012MNRAS.425..460M, 2017ApJ...835..165B, 2017MNRAS.464..946S, 2019ApJ...877...87Z, 2022Natur.603..237S, 2022arXiv220407282G}. The locations of the resulting sBBHs in its host galaxies from different mechanisms may be different, for example, the dynamical channel produces sBBHs in globular clusters (possibly outer regions of its host galaxy) or galactic nuclei, the AGN-MBH channel produces sBBHs only in galactic nuclei, while the EMBS channel produces sBBHs in areas across the whole host. Therefore, identifying the location of individual mergers of sBBHs detected by GW observatories will provide important information to distinguish different formation mechanisms for sBBHs.

The detection of GW signals itself can provide an estimate of the source sky location with an error typical of tens square degrees or larger, which is not sufficient for identifying the hosts. Searching for EM counterparts of GW sources is currently the only way to find the hosts. For example, the kilonova signal from GW170817 clearly indicates this GW source is located at the outer skirt of the elliptical galaxy NGC\,4993 \citep{2017ApJ...848L..12A}. However, EM counterpart searches for the GW detected sBBH mergers did not firmly find the host galaxy of any event \citep[e.g.,][but a possible candidate for GW190521, \citealt{2020PhRvL.124y1102G}]{2019PhRvX...9c1040A, 2020arXiv201014527A,2021arXiv211103606T,2021arXiv211103634T}. The main reason might be that sBBH mergers do not have EM counterparts or only have extremely faint ones (if any). Therefore, it is demanding to find new ways to identify the host galaxies of sBBH mergers. 

GWs from sBBH mergers can also be gravitational lensed by intervening galaxies \citep[e.g.,][]{1996PhRvL..77.2875W, 1998PhRvL..80.1138N, 2018PhRvD..98j4029D} and such lensed GW events are expected to be detected by Einstein Telescope (ET) and Cosmic Explorer (CE), respectively, with significant rate  \citep[e.g.,][]{ 2014JCAP...10..080B, Piorkowska:2013eww, 2015JCAP...12..006D, 2018MNRAS.476.2220L, 2019ApJ...874..139Y, PhysRevD.103.104055, 2021MNRAS.501.2451M, 2021ApJ...921..154W,2022MNRAS.509.3772Y}. The time delay(s) between different images can be precisely measured, which enables the lensed GW sources to be unique probes to constrain cosmological parameters \citep[e.g.,][]{2017NatCo...8.1148L, 2019ApJ...873...37L,2020MNRAS.498.3395H}, provided its positions in (or associated with) the host galaxies can be determined. 

The positions of the lensed sBBH mergers in its host galaxies may be determined by discovering the lensed EM signal of the hosts, because most lensed sBBH mergers are not likely to have (detectable) EM counterparts. Assuming that a lensed sBBH merger is associated with one of the lensed galaxies discovered in the sky area of the event by large sky surveys, the host galaxy of the lensed events may be identifiable by comparing the time-delay(s) measured from the GW signals and those inferred from the lensed galaxies \citep{2020MNRAS.497..204Y}.
Furthermore, the exact location of the GW event in its host galaxy may be also determined by combining both the lensed host galaxy and GW signals \citep[][]{2020MNRAS.498.3395H}. However, the assumption is not always satisfied as the lensed host galaxy may be too faint to be detectable by those surveys or its brightest part may be misaligned with the GW lens and thus does not have significant magnification or distortion to be identified as a lensed galaxy. Recently, \citet[][]{2022arXiv220408732W} first investigated about the probability of detecting the lensed host galaxy of a lensed GW event, specifically for the Euclid. They found that only a fraction of $\sim 20\%$ of the lensed hosts are detectable.

In principle, the locations of the sBBH mergers in its host galaxies can also be used to constrain its formation mechanisms as different formation channels can lead to significant different spatial distributions of these events in its hosts. In this paper, we investigate the detectability of the host galaxies of those lensed sBBH GW events generated via different sBBH formation channels by future (survey) telescopes, such as the Chinese Space Survey Telescope (CSST), Euclid, and the Nancy Grace Roman Telescope (RST, formerly WFIRST), etc. We also futher estimate the distributions of the lensed sBBH mergers in its hosts with detectable signatures and demonstrate that is possible to distinguish different sBBH formation channels by using this distribution. 

This paper is organized as follows. In Section~\ref{sec:method}, we briefly introduce a simple method to calculate the conditional probability of those lensed GW events that can have its host galaxies being identified as the lensed ones by a galaxy survey. In Section~\ref{sec:results}, we present our main results. Discussions and conclusions are given in Section~ \ref{sec:con}. Throughout the paper, we adopt the cosmological parameters as $(h_0,\Omega_{\rm m},\Omega_\Lambda)=(0.68,0.31,0.69)$ \citep{Aghanim2020}.

\section{method}
\label{sec:method}

\subsection{Lenses}

The galaxy-galaxy lensing cross section are mostly dominated by elliptical galaxies, which could be approximated by isothermal mass or elliptical power law (EPL) model \citep[e.g.,][]{Oguri_2010,2017MNRAS.465.4895W}. Previous and current observations show that most strongly lensed systems have double or qudra images. Therefore, it is reasonable to apply the singular isothermal ellipsoid profile (SIE) as the lens model. The lensing probabilities distribution are dependent on the foreground galaxy redshift $z_{\rm l}$ and velocity dispersion $\sigma_{\rm v}$ \citep[e.g.,][]{Collett_2015,2020MNRAS.497..204Y}:
\begin{equation}
d\tau = S_{\rm cr}(\sigma_{\rm v},z_{\rm l},z_{\rm s}) \frac{dn(\sigma_{\rm v},z_{\rm l})}{d\sigma_{\rm v}} \frac{dV({z_{\rm l}})}{dz_{\rm l}}d\sigma_{\rm v}dz_{\rm l},
\label{eq:tau}
\end{equation}
where $S_{\rm cr}\equiv \pi \theta_{\rm E}^{2}$ is the lensing cross-section with $\theta_{\rm E}$ denoting the Einstein radius, the velocity distribution function (VDF) $dn(\sigma_{\rm v},z_{\rm l})/d\sigma_{\rm v}$ is the comoving number density of the galaxy lenses with velocity dispersion in the range from $\sigma_{\rm v}$ to $\sigma_{\rm v}+d\sigma_{\rm v}$, and $dV(z_{\rm l})$ is the comoving volume within redshift range from $z_{\rm l}$ to $z_{\rm l}+dz_{\rm l}$. 

The VDF of the foreground galaxy lenses may evolve with redshift. However, we ignore this redshift evolution and assume the VDF at any redshift is the same as that in the local universe given by \citep[e.g.,][]{2007ApJ...658..884C, Piorkowska:2013eww} 
\begin{equation}
\frac{dn(\sigma_{\rm v},z_{l})}{d\ln \sigma_{v}}=n_{0} \frac{\beta}{\Gamma(\alpha/\beta)}
\left(\frac{\sigma_{\rm v}}{\sigma_{\rm v0}}\right)^{\alpha}\exp{\left[-\left(\frac{\sigma_{\rm v}}{\sigma_{\rm v0}}\right)^{\beta}\right]},
\label{eq:lens}
\end{equation}
where $\sigma_{\rm v0}$ is the characteristic velocity dispersion, $\alpha$ is the low-velocity power-law index, $\beta$ is the high-velocity exponential cutoff index, ${\Gamma(\alpha/\beta)}$ is the Gamma function, and $(n_{0}, \sigma_{\rm v0}, \alpha, \beta) = (0.008h^{3} {\rm Mpc}^{-3}, 161{\rm km\,s^{-1}}, 2.32, 2.67)$.

One may note that the actual VDF does evolve with redshift \citep[e.g., see][]{2022arXiv220106761Y}. We do not consider this evolution because the VDF does not vary significantly at $z_{\rm l} \lesssim 1.5$ \citep{2007ApJ...658..884C,2022arXiv220106761Y} and the VDF at $z_{\rm l}>1.5$ is still not available in the literature due to observational limitation. The assumption of a non-evolving VDF here may not lead to significant effect on the lensing rates estimated in this paper as most lens galaxies are at redshift $z_{\rm l} \sim 0-1$. 

\subsection{GW events: redshift distribution}
\label{merger}

The distribution of the sBBH GW events and its host galaxies can be described by the merger rate density as
\begin{equation}
\frac{dN}{dz_{\rm s} dt d\mathcal{M}_{\rm c}}=\frac{\boldsymbol{R}(z_{\rm s},\mathcal{M}_{\rm c})}{1+z_{\rm s}}\frac{dV(z_{\rm s})}{dz_{\rm s}}
\label{source}
\end{equation}
where $\mathcal{M}_{\rm c}$ is the chirp mass, $\boldsymbol{R}(z_{\rm s},\mathcal{M}_{\rm c})$ is the merger rate density with the chirp mass in the range from $\mathcal{M}_{\rm c}$ to at $\mathcal{M}_{\rm c}+ d\mathcal{M_{\rm c}}$ at redshift $z_{\rm s}$, and the factor $1/(1+z_{\rm s})$ accounts for the time dilation. The predicted merger rate density evolution are different for different sBBH formation channels. We adopt the following simple models to estimate $\boldsymbol{R}(z_{\rm s},\mathcal{M}_{\rm c})$ generated from different sBBH formation channels. 

$\textbf{EMBS channel:}$ The merger rate density for sBBHs formed via the EMBS channel is simply estimated as
\begin{eqnarray}
\boldsymbol{R_{\rm B}}\left(z_{\rm s}, \mathcal{M}_{\rm c}\right) &= &\iiint d \tau_{\mathrm{d}} d q dm_1 f_{\rm eff}  R_{\text {birth }}(m_{1}, \mathbf{z^{\prime}})
\nonumber \\
& & \times P_{\tau_{\rm d}}\left(\tau_{\mathrm{d}}\right) P_{q}(q)\delta\left(m_1-s^{-1}(\mathcal{M}_c,q)\right), \nonumber \\
\label{eq:Ptaud}
\end{eqnarray}
and 
\begin{eqnarray}
R_{\mathrm{birth}}\left(m_{1}, \mathbf{z^{\prime}}\right) & = &\iint d m_{\ast} d Z \dot{\psi}\left(Z ; \mathbf{z^{\prime}}\right) \nonumber \\ 
& &\times \phi\left(m_{\ast}\right) \delta\left(m_{\ast}-g^{-1}\left(m_{1}, Z\right)\right).
%
%
\end{eqnarray}
Here $\mathcal{M}_c=s(m_1,q)$ is the chirp mass as a function of the primary mass $m_1$ and mass ratio $q$, $f_{\rm eff}$ denotes the formation efficiency of sBBHs, which can be calibrated by the local sBBH merger density given by LIGO/Virgo observations, $\dot{\psi}(Z; \mathbf{z^{\prime}})$ is the cosmic star formation rate density (SFR) with metallicity $Z$ at formation redshift $\mathbf{z^{\prime}}$, $\phi(m_{\ast})$ is the initial mass function (IMF), and $m_1=g(m_{\ast,Z})$ is the relationship between the initial zero age main sequence star mass and the remnant mass given by \citet{2015MNRAS.451.4086S}. In the above Equation~\eqref{eq:Ptaud}, {$\mathbf{\tau_{\rm d}(z^{\prime})=\int_{z_{s}}^{{z^{\prime}}}|\frac{dt}{dz}|dz}$} denotes the time delay of a sBBH merger from its progenitor binary star formation time, and its probability distribution is assumed to be $P_{\tau_{\rm d}}(\tau_{\rm d}) \propto \tau_{\rm d}^{-1}$ \citep[e.g.,][]{2016Natur.534..512B,2016MNRAS.461.3877D}. The minimum and maximum values of $\tau_{\rm d}$ are set as $50$\,Myr and the Hubble time, respectively. The distribution of mass
ratio $q = m_1/m_2$ is assumed to be proportional to $q$ in the
range from 0.5 to 1 \citep[see][]{2016Natur.534..512B}.

We assume in this paper that $\dot{\psi}(Z ;\mathbf{z^{\prime}})$ can be separated to two independent functions, one is the total SFR $\dot{\chi}(\mathbf{z^{\prime}})$ at the formation redshift $\mathbf{z^{\prime}}$ and the other is metallicity distribution of those stars [$P(Z|\mathbf{z^{\prime}})$] at that redshift. We adopt the total SFR obtained from observations \citep[see][]{2014ARA&A..52..415M} as
\begin{equation}
\dot{\chi}(\mathbf{z^{\prime}})=0.015 \frac{(1+\mathbf{z^{\prime}})^{2.7}}{1+[(1+\mathbf{z^{\prime}}) / 2.9]^{5.6}} M_{\odot} \mathrm{Mpc}^{-3} \mathrm{yr}^{-1},
\end{equation}
and a log-normal metallicity distribution $P(Z|\mathbf{z^{\prime}})$ with a mean given by \citep[see][]{2016Natur.534..512B}
\begin{eqnarray}
\log \left[Z_{\text {mean }}(\mathbf{z^{\prime}})\right]& = &0.5+\log \left(\frac{y(1-R)}{\rho_{\mathrm{b}}}\right . \nonumber\\
 & & \left . \times \int_{\mathbf{z^{\prime}}}^{20} \frac{97.8 \times 10^{10} \dot{\phi}\left(z\right)}{H_{0} E\left(z\right)\left(1+z\right)} d z\right),
\end{eqnarray}
where the return fraction $R = 0.27$,  the net metal yield $y = 0.019$, the baryon density $\rho_{\rm b} = 2.55\times10^{11}\Omega_{\rm b}h_{0}^2M_{\odot}\rm Mpc^{-3}$ with $\Omega_{\rm b}=0.045$, $H_0$ is the Hubble constant, and $E(z)=\sqrt{\Omega_{\rm m}(1+z)^3+\Omega_{\Lambda}}$. The scatter of this log-normal metallicity distribution is $0.5$\,dex. 

$\textbf{Dynamical channel:}$ The merger rate density for sBBHs via the dynamical channel in dense (globular) clusters can be estimated as in \citet{2021MNRAS.500.1421Z} by using both dynamical simulations on the formation of sBBHs and simple descriptions on the formation and evolution of globular clusters. That is
\begin{equation}
\begin{aligned}
& \boldsymbol{R_{\rm D}}(z_{\rm s},\mathcal{M}_{\rm c})= \left.\iiint \frac{\dot{M}_{\mathrm{GC}}}{d \log _{10} M_{\mathrm{Halo}}}\right|_{z(\tau)} \frac{1}{\left\langle M_{\mathrm{GC}}\right\rangle} P\left(M_{\mathrm{GC}}\right) \\
& \times R\left(r_{\mathrm{v}}, M_{\mathrm{GC}}, \tau-t(z_{\rm s})\right) P(\mathcal{M}_{\rm c}) d M_{\mathrm{Halo}} d M_{\mathrm{GC}} d \tau,
\end{aligned}
\end{equation}
where $\frac{\dot{M}_{\mathrm{GC}}}{d \log _{10} M_{\mathrm{Halo}}}$ is the comoving SFR in globular clusters per galaxies of a given halo mass $M_{\rm Halo}$ at given redshift $z(\tau)$ (or a given formation time $\tau$ ), $P(M_{\rm GC})$ is the cluster initial mass function, ${\left\langle M_{\mathrm{GC}}\right\rangle}$ is the mean initial mass of a globular cluster, $R(r_{\rm v}, M_{\rm GC}, t)$ is the merger rate of sBBHs in a globular cluster with initial virial radius $r_{\rm v}$ and mass $M_{\rm GC}$ at time $t(z_{\rm s})$, and $P(\mathcal{M}_{\rm c})$ is the chirp mass distribution of sBBH mergers produced by the dynamical interactions in globular cluster given in \citet{2018ApJ...866L...5R}. We adopt the specific form for $\frac{\dot{M}_{\mathrm{GC}}}{d \log _{10} M_{\mathrm{Halo}}}$ and $R(r_{\rm v}, M_{\rm GC}, t)$ obtained by \citet{2018ApJ...866L...5R}, which assumes $50\%$ of clusters form with $r_{\mathrm{v}} = 1$\,pc and $50\%$ form with $r_{\mathrm{v}} = 2$\,pc.  The total rate is the summation of those from the two $v_{\rm v}$ cases.

More detailed descriptions about the estimates of the cosmic evolution of sBBH merger rate density via the above two channels can be found in \citet{2017cao} and \citet{2021MNRAS.500.1421Z}, respectively.

$\textbf{AGN-MBH channel:}$ The formation of sBBHs via the AGN-MBH channel is intensively discussed recently and proposed as the origin of some LIGO/Virgo GW events, such as GW190521 \citep[e.g.,][]{2021ApJ...907L..20T, 2022Natur.603..237S}. Several papers have estimated the local merger rate density for sBBHs produced by AGN/MBH channel, in a wide range from $0.02$ to $60 \rm Gpc^{-3}yr^{-1}$ \citep[e.g.,][]{2017MNRAS.464..946S,2019MNRAS.488...47F,2020ApJ...898...25T}, which suggests that the AGN/MBH channel could be significant or even dominant comparing with other channels.
However, estimation for the merger rate density evolution of sBBHs via this channel is still not available in the literature. Nevertheless, it is plausible to assume that the merger rate density $\boldsymbol{R_{\rm A}}(z_{\rm s})$ via the AGN-MBH channel at redshift $z_{\rm s}$ is proportional to the total rate of accreted mass on to MBHs at that redshift. Therefore, the merger rate density can be approximated as
\begin{equation}
\label{qso}
\boldsymbol{R_{\rm A}}(z_{\rm s})\propto \dot{\rho}_{\bullet}^{\mathrm{QSO}}(z_{\rm s}) 
= \int_{0}^{\infty} \frac{(1-\epsilon) L_{\mathrm{bol}}}{\epsilon c^{2}} \Psi(L_{\rm bol}, z) \mathrm{d}L_{\rm bol},
\end{equation}
where $\Psi(L_{\rm bol}, z)$ is the bolometric luminosity function of QSOs, $\epsilon$ is the radiative efficiency of the accretion processes and assumed to be the canonical value $0.1$ \citep[e.g.,][]{2002MNRAS.335..965Y, 2004ApJ...602..603Y, 2004MNRAS.351..169M, 2007ApJ...654..731H, 2008MNRAS.383..277K, 2022MNRAS.509.3488I}. Since equation~\eqref{qso} contains no chirp mass information, which is requisite for the calculation of the GW signal-to-noise ratio (SNR), i.e., $\varrho_{\rm GW}$, we simply adopt the same marginal chirp-mass distribution as that from the dynamical channel. We will show later that, profited by the high sensitivity of the 3rd generation GW detectors, different chirp mass distribution adopted here will not affect our main conclusions much. 

\begin{figure}
\includegraphics[width=1.0\columnwidth]{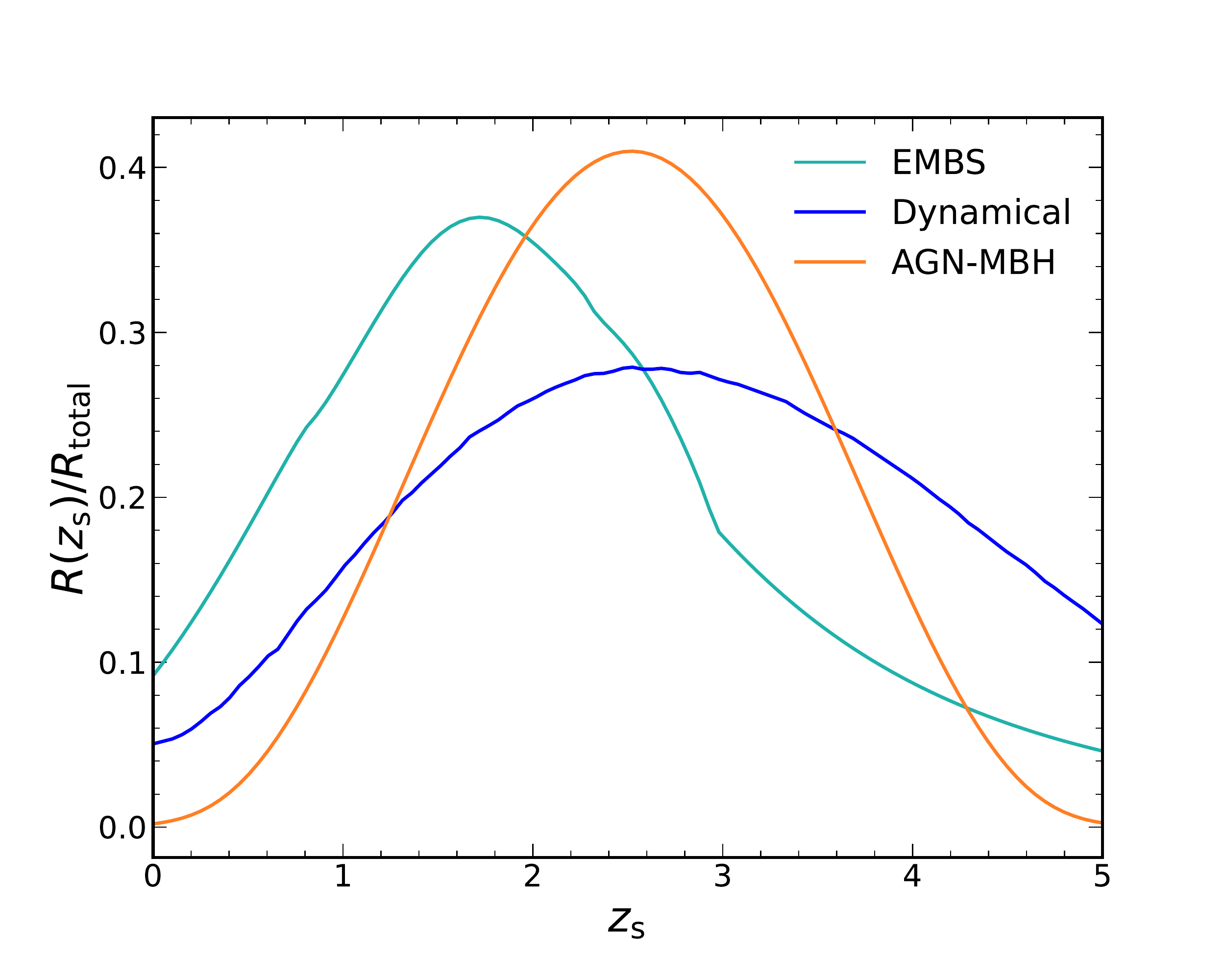}
\centering
\caption{The normalized merger rate density evolution as a function of redshift. Here $\boldsymbol{R}(z_{\rm s})$ represent the merger rate density at redshift $z_{\rm s}$ and $\boldsymbol{R}_{\rm total}$ is the integration of the merger rate density over  redshift. Cyan, blue, and orange lines represent the results obtained by assuming that all sBBHs are produced by the EMBS, dynamical, and AGN-MBH channels, respectively. 
}
\label{fig:fig1}
\end{figure}

Figure~\ref{fig:fig1} shows the estimates for merger rate density evolution obtained for the above simple approximations after marginalizing the sBBH's chirp-mass, i.e., 
\begin{equation}
\boldsymbol{R}(z_{\rm s})=\int \boldsymbol{R}(z_{\rm s},\mathcal{M}_{\rm c})d\mathcal{M}_{\rm c}.
\end{equation}
In this figure, the cyan, blue, and orange lines represent the merger rate density evolution obtained from the EMBS, dynamical, and AGN-MBH formation channels, respectively, which are different from each other. Apparently, the merger rate density from the EMBS channel peaks at $z_{\rm s}\sim 1.5$, while those from the dynamical and AGN-MBH channels peak at higher redshifts, i.e.,  $z_{\rm s}\sim 2.5$; the merger rate density from the AGN-MBH decreases more rapidly at both the low-redshift and high-redshift ends, comparing with those from the other two channels.

\subsection{GW events: spatial distribution in the host}
\label{spatial}

The spatial distributions of the resulting sBBH mergers in its host galaxies may be different for different sBBH formation channels. In order to generate mock GW lensed events and the images of its lensed host galaxies, we adopt some simple approximations for the spatial distribution of sBBH mergers in their host galaxies for each formation channel as detailed below. 

\begin{itemize}
\item sBBH mergers formed by the EMBS channel can occur anywhere in its host galaxy. The probability distribution for the location of these sBBH mergers in its host should depend on the assembly history of the host and the formation history of stars in it, which is rather complicated. However, this spatial distribution may more or less follow the mass distribution of stars, and thus the luminosity distribution of the host, if the mass-to-light ratio is close to a constant. In the following calculations, we simply assume that the probability distribution of the location of a sBBH merger produced via the EMBS channel is proportional to the stellar mass and their distribution of its host.  

\item sBBH mergers formed via the dynamical channel are likely to locate in the GCs of its host galaxies. One direct consideration would be assuming that the spatial distribution of these mergers follow the distributions of GCs in its hosts. Although the number of sBBH mergers produced via this channel depends on the properties of GCs, the assumption is still reasonable, provided that the physical properties of GCs statistical do not depend on its location \citep[e.g.,][]{2000ApJ...528L..17P,2016ApJ...824L...8R,2017cao}. Therefore, we adopt this simple assumption below to set the spatial distribution of sBBHs in its hosts for our calculations.

\item sBBH mergers formed via the AGN-MBH channel locate at the center of its host galaxies. Comparing with the scale of its host galaxies, the spatial distribution of these sBBH mergers can be approximated as a $\delta$ function located right at the galactic centers.
\end{itemize}   

To obtain estimates on the number of lensed GW events that actually have distinguishable lensed signatures (such as significantly magnified arcs, rings, or so) from their host galaxies, it is necessary to have the information on the distributions of the spatial and physical properties of the hosts and the spatial distributions of the GW events in the hosts. Recent cosmological hydrodynamical simulations, such as Illustris TNG \citep[e.g.,][]{2018MNRAS.475..676S, 2018MNRAS.475..648P}, Eagle \citep[e.g.,][]{2015MNRAS.450.1937C, 2016A&C....15...72M}, etc., generated mock galaxies across the cosmic time. One can get the spatial distributions of stars and star formation rate directly from the simulations for each of these mock galaxies, while the distributions for the GCs in these mock galaxies are not available. Therefore, it is not straightforward to generate the location distribution for those lensed sBBH mergers via the dynamical channel in their hosts by using those simulations. Nevertheless, there is  evidence supporting that the formation of GCs are strongly related to the dark matter (DM) haloes so that GCs can be used as tracers of the detailed structural properties of the DM haloes of their host galaxies \citep[e.g.,][]{2017MNRAS.472L.104F, 2018MNRAS.477.3869H, 2022MNRAS.513.3925R}. Thus, for the purpose of this paper, we simply assume that for each sampled host, the spatial distribution of GCs within it follows the same distribution of the DM halo density profile, i.e., the Navarro-Frenk-White (NFW) profile, projected over the line of sight as
\begin{equation}
\rho_{\rm GC}(R,z_{\rm s})\propto  \int^{\infty}_{-\infty}\frac{dl}{\left(\sqrt{l^2+R^2} / r_{\mathrm{s}}\right)\left(1+\sqrt{l^2+R^2} / r_{\mathrm{s}}\right)^{2}}
\end{equation}
where $l$ is the line of sight length, $r_{\rm s}$ is the scale radius related with virial radius $r_{\rm s}=r_{\rm v}/c$, and $c$ is the concentration parameter, which can be expressed as the function of halo virial mass $M_{200}$ and redshift $z_{\rm s}$ \citep[e.g.,][]{1996ApJ...462..563N, 1997ApJ...490..493N, 2003ApJ...597L...9Z, 2005gfe..book.....S, 2009ApJ...707..354Z, 2014MNRAS.441..378L, 2020Natur.585...39W}. Here we adopt the fitting formula given by \citet[][]{2020Natur.585...39W}, i.e.,
\begin{equation}
\label{eq:c_vWang}
c=\frac{1}{1+z_{\rm s}}\exp \left[c_{6} \left(\frac{M_{\mathrm{fs}}}{M_{200}}\right)^{\frac{1}{3}}\right] \times \sum_{i=0}^{5} c_{\mathrm{i}}\left[\ln \frac{M_{200}}{h_0^{-1} M_{\odot}}\right]^{i},
\end{equation}
where $c_{i}=\left[27.112,-0.381,-1.853 \times 10^{-3},-4.141 \times\right.$ 
$\left.10^{-4},-4.334 \times 10^{-6}, 3.208 \times 10^{-7},-0.529\right]$ for $i \in\{0, \ldots, 6\}$ and the free-streaming mass scale $M_{\mathrm{fs}}=7.3 \times 10^{-6} M_{\odot}$.

\subsection{Samplings and Criteria}
\label{criteria}

Considering the relative positions of the lens and source, we use the code \textbf{LensPop} \citep[e.g.,][]{Collett_2015} to generate $10^7$ pair foreground and back ground galaxies with redshift in the range from $0$ to $5$, according to  equations~\eqref{eq:tau} and \eqref{eq:lens}. The angular positions are uniformly and randomly sampled in the sky. Intrinsic properties including apparent magnitude $m_{r,i,z}$ in $\rm r, i, z$ band, halo viral mass $M_{200}$, stellar mass $M_{\ast}$, ellipticity $q_{\rm s}$, effective radius $R_{\rm e}$ are also sampled at same time with consistency approximation in \citet{Collett_2015} using the sky catalogs simulated by the LSST collaboration \citep[e.g.,][]{2010SPIE.7738E..1OC}. 

Then, we randomly sample different numbers of GW-emitting sBBH mergers to the hosts according to their stellar mass $M_{\ast}$ by the spatial distribution of the merger event as described in Section~\ref{spatial} for different formation channels. The orientation angles, i.e., ( $\theta_s$, $\phi_s$, $i$, $\psi$)\footnote{Here $\theta_{\rm s}$ and $\phi_{\rm s}$ are the declination (Dec) and right ascension (RA) of the GW source in the celestial coordinate system, while $i$ and $\psi$ give the source’s orientation with respect to the detector.} are all uniformly and randomly sampled in the sky.

We use the standard package pyCBC \citep[][]{2019PASP..131b4503B} to generate the GW waveform for each sBBH, by adopting the phenomenological model \textbf{IMRPhenomPv3} proposed by \citet{2019PhRvD.100b4059K}, in which the dynamics of precessing binary black holes with two-spin effects are all considered. The total strain $h(f)$ received by a GW detector is
\begin{equation}
h(f)=F_{+}(f)h_{+}(f)+F_{\times}(f)h_{\times}(f),
\end{equation}
where $F_{+}$ and $F_{\times}$ is the detector's pattern function, of which the explicit expressions in the time-domain (i.e., $\Tilde{F}_{+,\times}(t)$)  are periodic functions of time with a period equal to one sidereal day, due to the diurnal motion of the Earth \citep[e.g.,][]{PhysRevD.58.063001,PhysRevD.81.062003,2018PhRvD..97f4031Z}.

Here we define the whitened GW data sets of a GW network composed of $n$ detectors (e.g., $n=1$ for a single detector) as
\begin{equation}
\hat{\mathbf{d}}(f)=\left(\frac{A_1(f)h_1(f)}{\sqrt{S_{1}(f)}},\frac{A_2(f)h_2(f)}{\sqrt{S_{2}(f)}}, ..., \frac{A_{n}(f)h_{n}(f)}{\sqrt{S_{n}(f)}}\right),
\end{equation}
where $A_{n}=e^{-2\pi i f((\hat{r}_{n}-\hat{r}_{1})\cdot \hat{n}_{\rm GW})}$ is the phase transfer function, $\hat{r}_{n}$ is the location of the n-th detector, $\hat{n}_{\rm GW}$ is the unit direction vector of the GW source, and $S_{n}$ denotes the one-sided power spectrum of the corresponding $n$-th GW detector \citep[][]{2010PhRvD..81h2001W}. 
Then the optimal squared SNR is given by
\begin{equation}
\varrho_{GW}^2=\left\langle\hat{\mathbf{d}}(f)|\hat{\mathbf{d}}(f)\right\rangle,
\label{eq:SNR}
\end{equation}
where the angular bracket denotes an inner product. For any two vector functions $\hat{\mathbf{a}}(f)$ and $\hat{\mathbf{b}}(f)$, this inner product is defined as
\begin{equation}
\langle \hat{\mathbf{a}}(f)| \hat{\mathbf{b}}(f)\rangle=2\sum_{j} \int_{f_{\text {min }}}^{f_{\text {max }}}\left\{a_j(f) b_j^{*}(f)+a_j^{*}(f) b_j(f)\right\} d f ,
\end{equation}
where $j$ denote for the $j$-th component of the vector, ${f_{\rm min}}$ and ${f_{\rm max}}$ are the lower and upper frequency limits of the GW waveforms.

The localization error for each sBBH GW source may be estimated according to the Fisher information matrix \citep[e.g.,][]{PhysRevD.46.5236,PhysRevLett.70.2984,2018PhRvD..97f4031Z,2022ApJ...932..102W,2022arXiv220702771I,2022arXiv220312586P}, $\Gamma_{jk}$, which is defined as 
\begin{equation}
\Gamma_{j k}=\left\langle\partial_{j} {\hat{ \mathbf{d}}(f)} \mid \partial_{k} \hat{\mathbf{d}}(f)\right\rangle,
\end{equation}
where $\partial_j$ and $\partial_k$ denote the partial derivative with respect to the $j$-th and $k$-th parameter, respectively. Once the Fisher matrix is determined, the covariance matrix of the location of a GW source in the celestial coordinates is given by
\begin{equation}
\rm Cov(\theta_{\rm s},\phi_{\rm s})=\Gamma^{-1}.
\label{eq:error}
\end{equation}
Considering the multiple magnified images of a lensed GW source, the total covariance matrix can be simply estimated by the summation of that for each image. With this total covariance matrix, we get the localization errors for a lensed sBBH merger in solid angle as
\begin{equation}
\Delta \Omega=2 \pi\left|\sin \theta_{\rm s}\right| \sqrt{\left\langle\Delta \theta_{\rm s}^{2}\right\rangle\left\langle\Delta \phi_{\rm s}^{2}\right\rangle-\left\langle\Delta \theta_{\rm s} \Delta \phi_{\rm s}\right\rangle^{2}},
\label{eq:omega}
\end{equation}
where $\Delta \theta_{\rm s}$ and $\Delta \phi_{\rm s}$ is the standard deviation obtained from the covariance matrix.  More detailed descriptions on the estimation of the GW localization precision for compact binary coalescences can be seen in \citet{2018PhRvD..97f4031Z} and \citet{2022arXiv220312586P}. 

We may estimate the detection probability of the lensing signatures for the host galaxies of lensed sBBH events as 
\begin{equation}
\label{bayes}
P({\rm H}|{\rm GW})=\frac{P({\rm GW}|{\rm H})P({\rm H})}{P({\rm GW}|{\rm H})P({\rm H})+P({\rm GW}|{\rm H^{\ast}})P({\rm H^{\ast}})},
\end{equation}
where ``H'' and ``GW'' denote those host galaxies that have identifiable lensing images and those detected GW events that are identified as lensed ones, ${\rm H^\ast}$ represents those lensed galaxies without identifiable lensing lensing images, and $P({\rm GW}|{\rm H})$, $P({\rm H})$, $P({\rm GW}|{\rm H^{\ast}})$, and $P({\rm H^{\ast}})$ represent the probability of those lensed host galaxies with identifiable lensing images that have identified lensed GW events, the probability of those lensed galaxies that have detectable lensing signatures, the probability for those lensed galaxies without detectable lensing signatures that have detected lensed GW event, and the probability of those lensed galaxies without detectable lensing signatures respectively. Samples contributing to each term are selected by the following criteria.
\begin{enumerate}
\item [1)] 
sBBHs should be within the caustic and the signal-to-noise ratio (SNR; $\varrho_{\rm GW}$) of the GWs from any one of the images should be larger than $8$, i.e., $\sqrt{\min(\{|\mu_{\rm m}|\})}\varrho_{\rm GW}>8$, where \{$|\mu_{\rm m}|$\} represents a list of the magnification factors for all the images. 
\item [2)] 
The center of the host galaxy should be within the caustic, thus the multiple lensed images could be identified, i.e., $x_{\rm s}^2+y_{\rm s}^2\lesssim \theta_{\rm E}^2$, where $\theta_{\rm E}$ is the Einstein radius.
\item [3)] The image and counter-image must be resolvable, i.e., $R_{\rm e}^2+s^2/2\lesssim \theta_{\rm E}^2$, where  $s$ is the seeing, e.g., $0.18^{\prime\prime}$ for the i-band of CSST \citep[e.g.,][]{2019ApJ...883..203G}, 
$0.23^{\prime\prime}$ for the VIS-band of Euclid \citep[e.g.,][]{2013LRR....16....6A} and $0.11^{\prime\prime}$ for the J-band of RST \citep[e.g.,][]{2013arXiv1305.5422S}.
\item [4)] The tangential shearing of the arcs should be detectable, i.e., ${\mu_{\rm tot}R_{\rm e}>s}$, where $\mu_{\rm tot}$ is the total magnification of the source. 
\item [5)] The lens signatures of the host galaxy should be detectable with sufficient $\rm SNR$ for identifying a strong lensing events ${\rm SNR} > 20$.
\end{enumerate}
 Noticed that, the above criteria is relatively stricter than those adopted in \citet[][]{2022arXiv220408732W}, even though the criteria  for identifying a lensed host themselves are quite flexible. As for the EM observations searching for the lensed hosts, we adopt the VIS-band magnitude of Euclid as $ m_{\rm VIS} = (m_{\rm r}+m_{\rm i}+m_{\rm z})/3$ \citep[see the same approximation in][]{Collett_2015}, and the J-band magnitude of RST as $ m_{\rm J}=m_{\rm z}-4.4(m_{\rm i}-m_{\rm z})$ \citep[see the approximation in][]{2020RNAAS...4..190W} for the mock host galaxies.   The power spectrum $S_n(f)$ of the 3rd generation GW detectors, including ET\footnote{ET-D design \citep[][]{Hild_2011} \href{http://www.et-gw.eu/}{http://www.et-gw.eu/} } and CE\footnote{Stage-2 phase \citep[][]{2019BAAS...51g..35R} \href{https://cosmicexplorer.org/}{https://cosmicexplorer.org/}}, are quite optimistic, and almost all sBBH mergers can be detected by these detectors. Thus the requirement of $\varrho_{\rm GW} \geq 8$ can be almost always satisfied. Therefore, we do not discuss any specific GW detector but instead propose a consistent estimation on the fraction of the lensed sBBH mergers that have identifiable lensed host galaxies, for all the 3rd generation
 GW detectors below. We also note here that only with the network of 3rd generation GW detectors, one may localize lensed sBBH mergers within a sky-area of $\mathbf{\sim \rm 5 deg^{2}}$. Therefore, the following estimations on the detection rates of such events are only applied for this powerful network.

\begin{figure*}
\includegraphics[width=0.68\columnwidth]{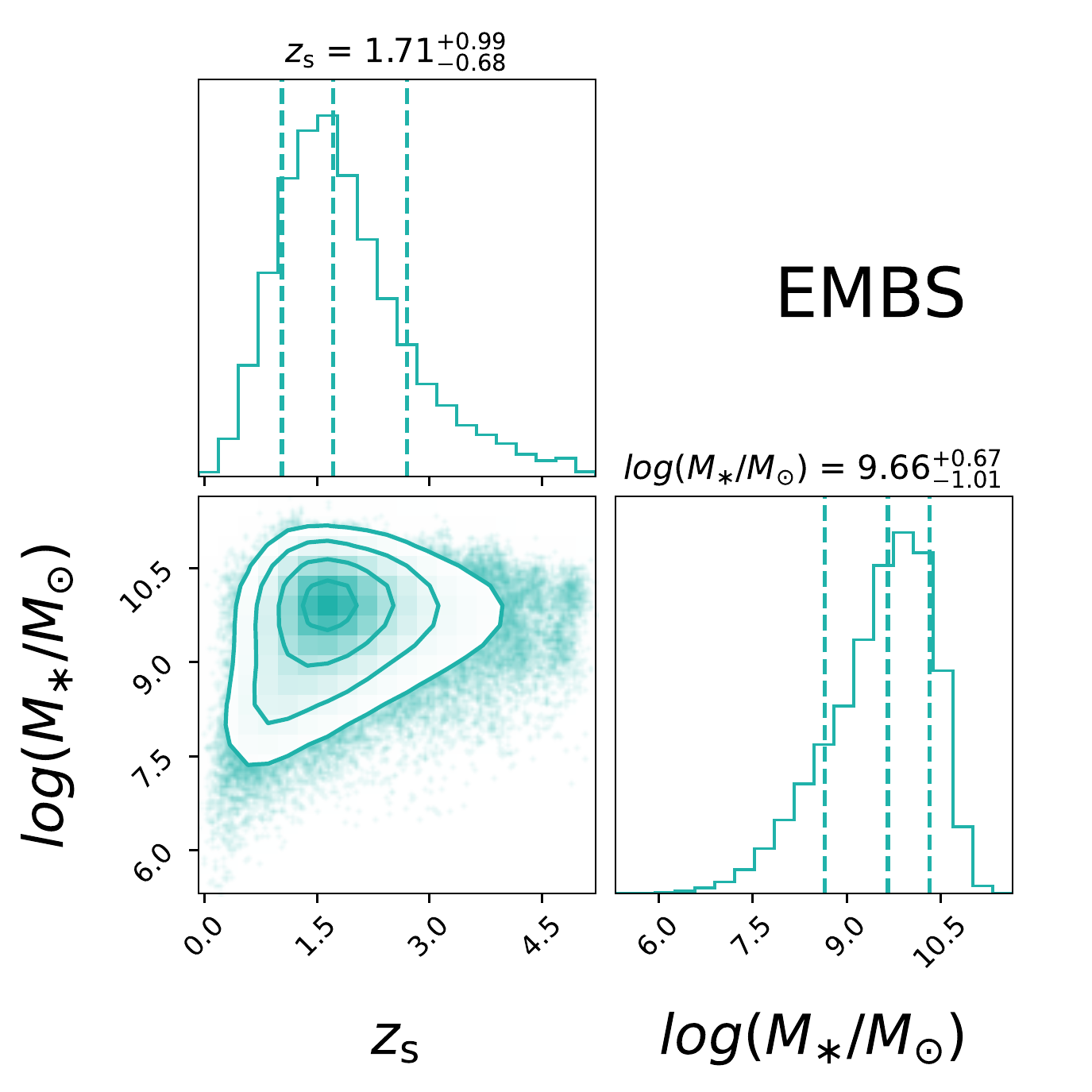}
\includegraphics[width=0.68\columnwidth]{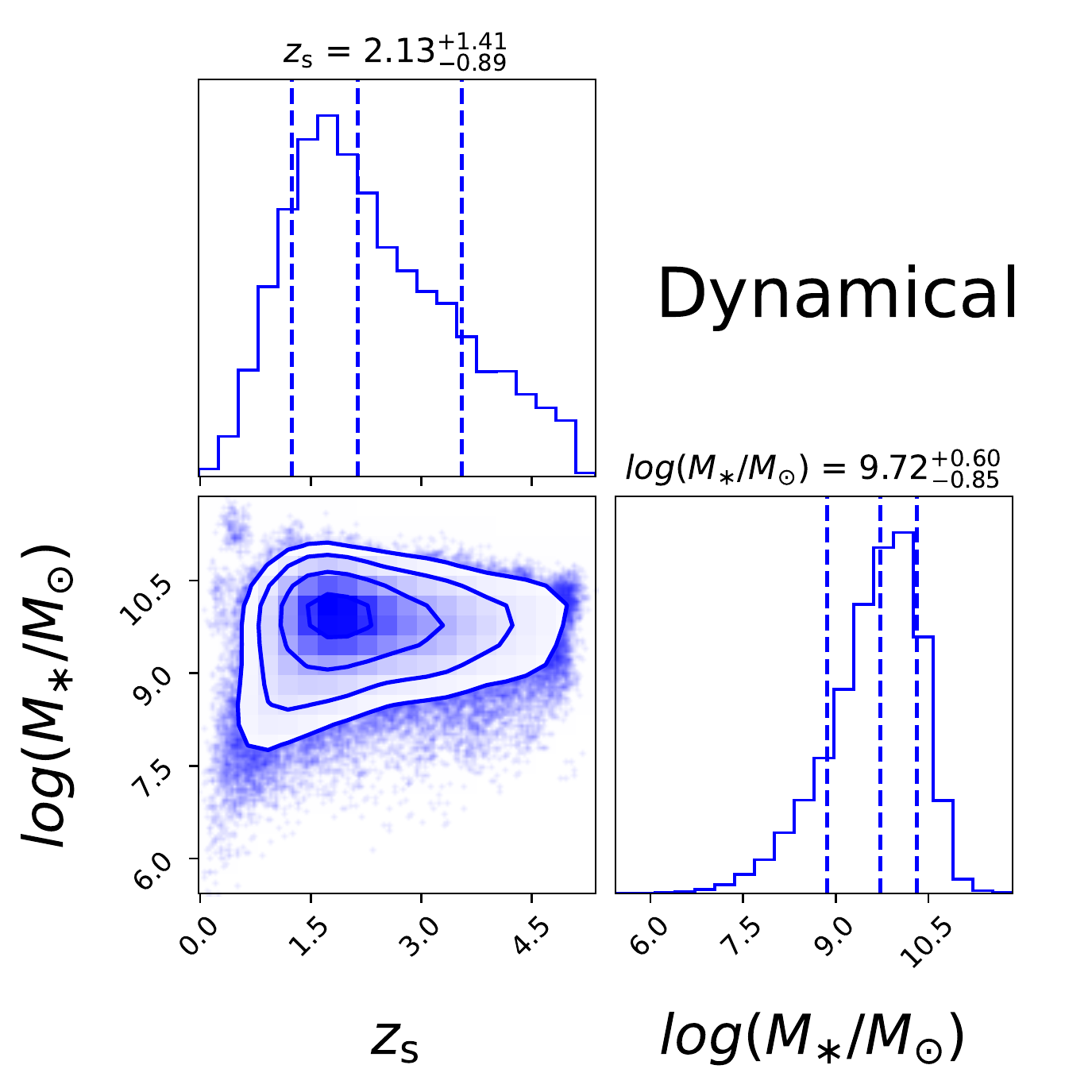}
\includegraphics[width=0.68\columnwidth]{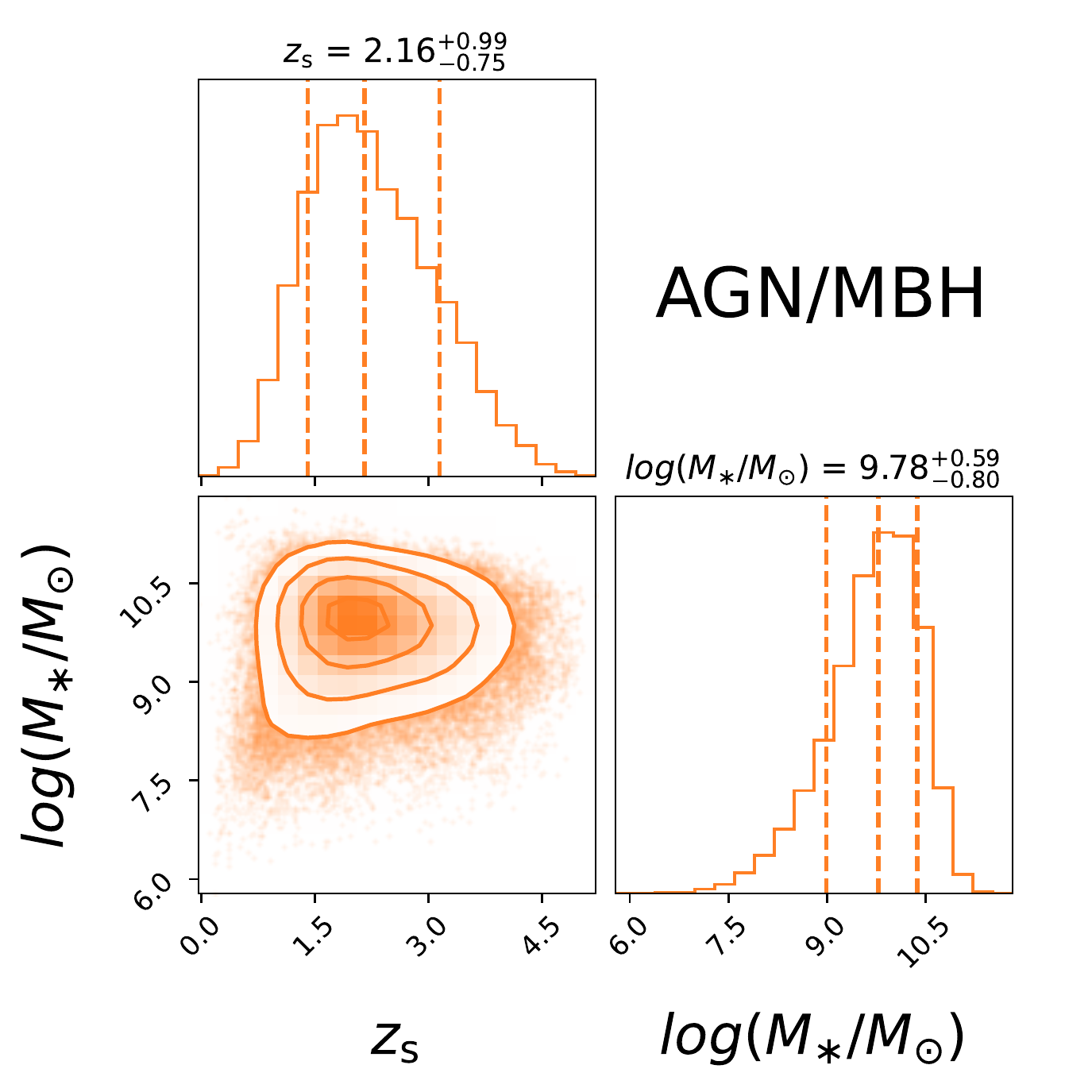}
\centering
\caption{
Redshift $z_{\rm s}$ and stellar mass $ \rm{log}(M_{\ast}/M_{\odot})$ distributions of the identifiable source galaxies in the simulated host samples for the lensed sBBH GW events, resulting from the EMBS (left), dynamical (mid), and AGN-MBH (right), respectively. The limiting magnitude for the survey to identify the lensing signatures of the host galaxies is set to be $m_{\rm i}=25$. The dashed lines and the solid contours indicate the $16\%$, $50\%$, and $84\%$ percentiles for each parameter.
}
\label{fig:embs_contour}
\end{figure*}

\section{Results}
\label{sec:results}

We obtain mock samples for both the lensed GW events and those with identifiable lensing signatures of its host galaxy for different sBBH formation models according to the settings and criteria described in section~\ref{sec:method}. We estimate the SNR and the localization error for each mock lensed GW event (Eqs.~\ref{eq:SNR} and \ref{eq:omega}) and then further calculate the probability of a lensed GW event that has identifiable lensing signatures of its host galaxy for different sBBH formation models.

Figure~\ref{fig:embs_contour} shows the redshift $z_{\rm s}$ and stellar mass $M_\ast$ distributions for the host galaxies of those sBBH mergers with both the lensed GW and host signals being identified. As seen from this figure, different sBBH formation channels may result in different $z_{\rm s}$ distribution but similar $M_\ast$ distribution. For example, the median redshifts of the identified hosts obtained by assuming the dynamical or AGN-MBH channels (with the median redshift at $2.13$ or $2.16$) are higher than that by assuming the EMBS channel ($1.71$), which is mainly due to the difference in the merger rate density evolution (see Fig.~\ref{fig:fig1}). The $M_\ast$ distributions obtained by assuming different formation channels have similar median and scatter. The $M_{\ast}$ distributions are all truncated at $\sim 10^{11}M_{\odot}$, partly caused by the lensing selection effects, i.e., massive galaxies are more likely to have a large effective radius, $R_{\rm e}$, which may violate the third criterion listed in in section~\ref{criteria}. Note here for demonstration purpose we only show the case with the limiting magnitude $m_{\rm i,lim}=25$\,mag for observations to identify the host lensing signatures. Choosing a different $m_{\rm lim}$, the results are almost the same. 

\begin{figure}
\includegraphics[width=1.1\columnwidth]{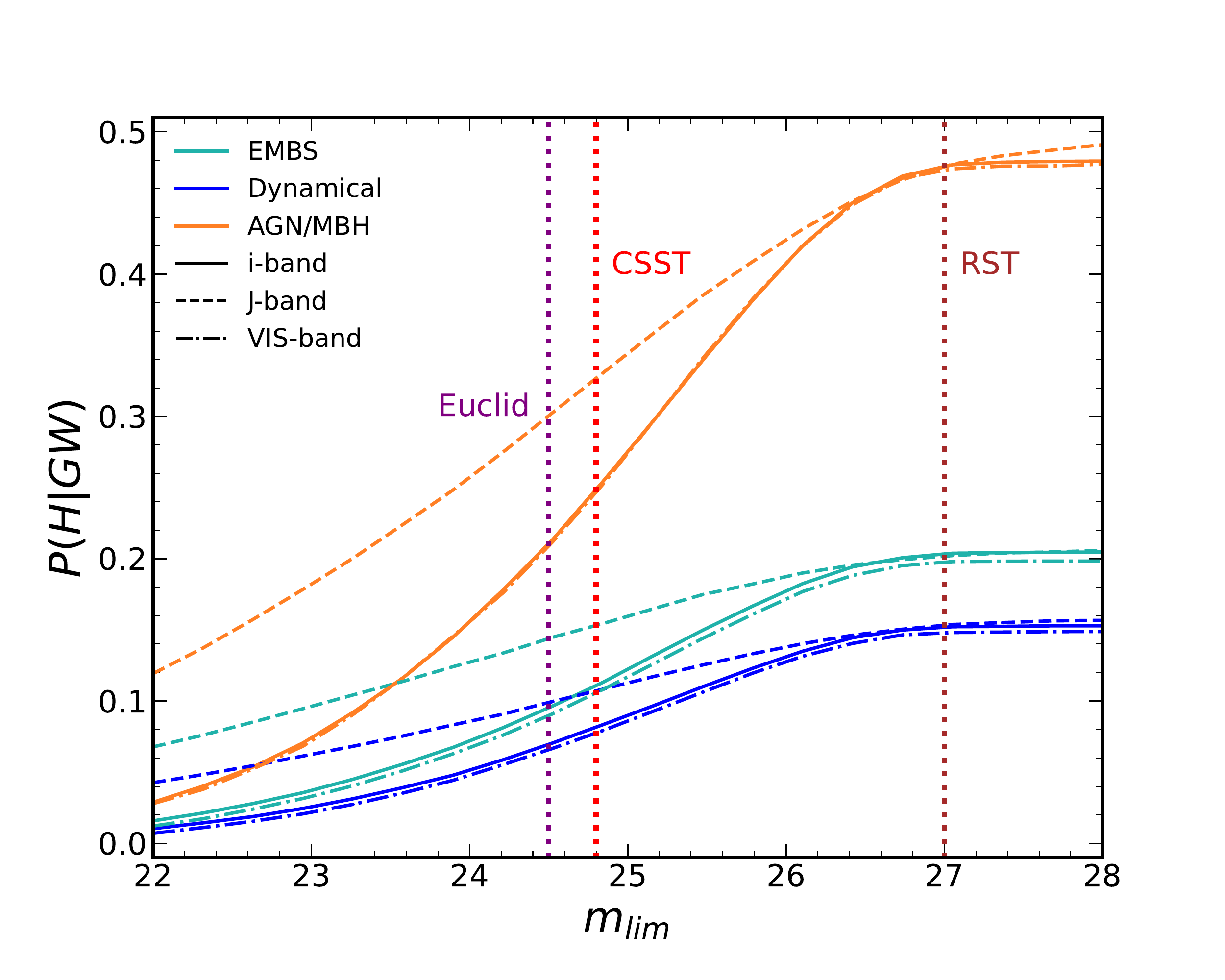}
\centering
\caption{The fraction of lensed GW sBBH  events $P({\rm H}|{\rm GW})$ that can have its host galaxies being identified as the lensed ones by a sky galaxy survey with the limiting magnitude of $m_{\rm lim}$. Solid, dashed, and dotted-dashed lines represent the i-band of CSST, the J-band of RST, and the VIS-band of Euclid, while cyan, blue, and orange lines show the results obtained by assuming that all the sBBH mergers are originated from the EMBS, dynamical, and AGN-MBH channel, respectively. The limiting magnitude of each telescope at the given band are also indicated by the purple, red, and brown dotted line.}
\label{fig:limit}
\end{figure}

Figure~\ref{fig:limit} shows the probability of a lensed sBBH GW event that has lensing signatures of its host galaxy (the conditional probability $P({\rm H}|{\rm GW})$) identifiable by a (survey) telescope with the limiting magnitude of $m_{\rm lim}$ for different sBBH formation models.  As seen from this figure, $P({\rm H}|{\rm GW})$ obtained for any sBBH formation channel increases with $m_{\rm lim}$ when $m_{\rm lim} \lesssim 26.5-27$ while becomes flat when  $m_{\rm lim} \gtrsim 27$. It is obvious that searching observations with larger $m_{\rm lim}$ can identify the lensing signatures of more lensed sBBH merger hosts and thus lead to larger $P({\rm H}|{\rm GW})$. When the searching observations are deep enough, all those lensed hosts that satisfy our criteria set in section~\ref{sec:method} can all be identified. However, there are a substantial fraction of the lensed hosts do not satisfy the criteria 2) and 3) in section~\ref{criteria} and cannot be identified, which is the primary reason for a flat $P({\rm H}|{\rm GW})$ at  $m_{\rm lim} \gtrsim 27$.

The probabilities $P({\rm H}|{\rm GW})$ resulting from different sBBH formation channels may show remarkable difference at  a given band with any given $m_{\rm lim}$. Assuming all the sBBH mergers are produced by the AGN-MBH channel leads to a substantially higher $P({\rm H}|{\rm GW})$ compared with that assuming either the EMBS or dynamical channel, and $P({\rm H}|{\rm GW})$ resulting from the case by assuming the dynamical channel is only slightly smaller than that from the EMBS channel. The reason is that all sBBHs from the AGN-MBH channel are located in the centers of its hosts and thus the criterion 2) in section~\ref{criteria} can be automatically satisfied, while a substantial fraction of the lensed hosts do not satisfy the criterion 2) in the cases assuming either the EMBS or dynamical channel. $P({\rm H}|{\rm GW})$ resulting from the dynamical channel is somewhat smaller than that from the EMBS channel because the dynamical channel leads to relative more sBBHs located at the outer skirt of its hosts, for which the lensing signatures of the hosts are harder to be identified. 

The probability $P({\rm H}|{\rm GW})$ is also dependent on which band is chosen for identifying the lensing signatures of the hosts. As seen from Figure~\ref{fig:limit}, choosing a redder band (e.g., J-band rather than i-band) leads a relatively larger $P({\rm H}|{\rm GW})$ at $M_{\rm lim}$. The reason is that those lensed hosts are usually located at high redshift thus more easier to be identified in a redder band than in a bluer band.

\begin{figure}
\includegraphics[width=1.0\columnwidth]{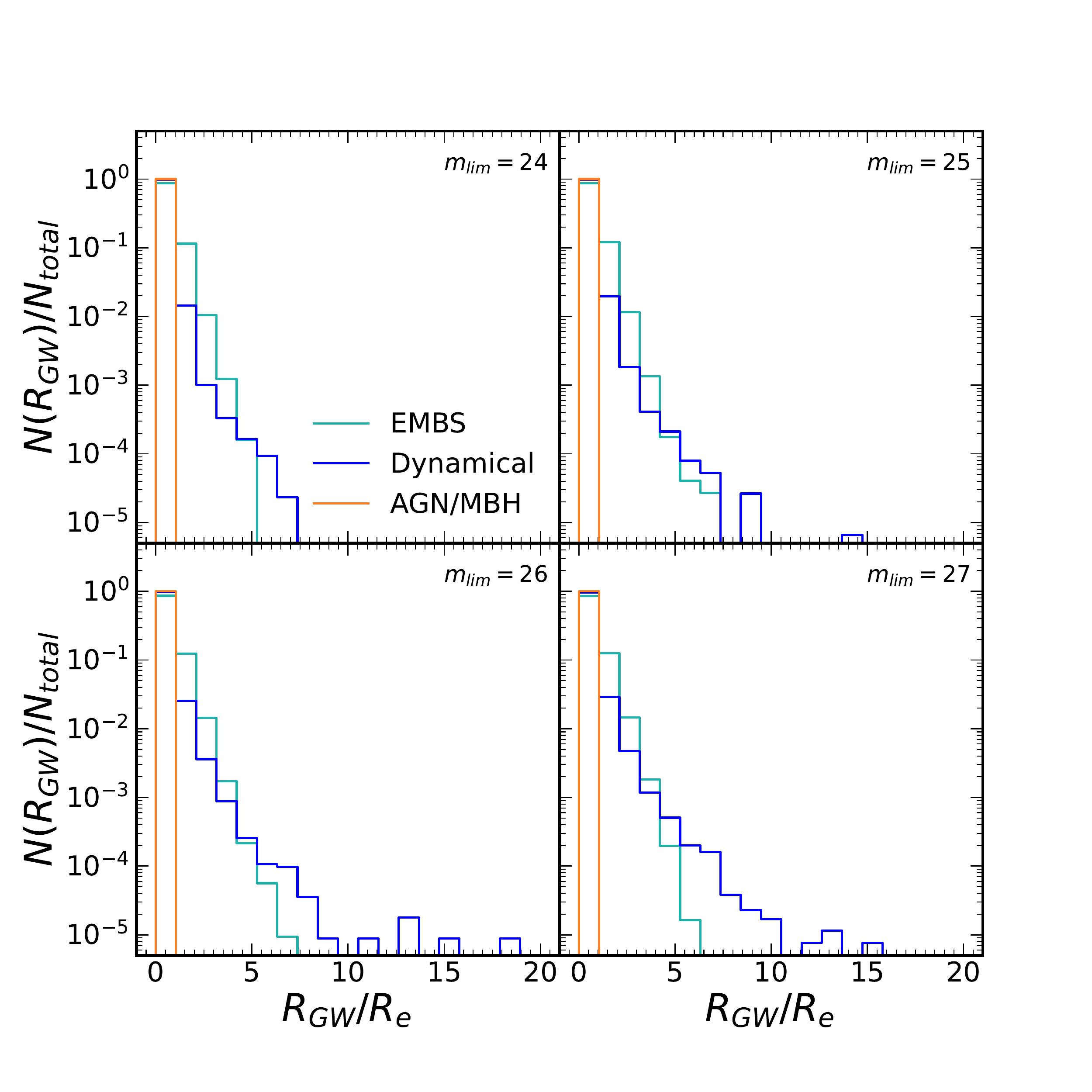}
\centering
\caption{Spatial distribution of the lensed GW sBBH events in the unit of the effective radius $R_{\rm e}$ of its host galaxies. The limit magnitude $m_{\rm lim}$ for the galaxy survey to identify the hosts as lensed ones is assumed to be $24$ (upper-left), $25$ (upper-right), $26$ (bottom-left) and $27$\,mag (bottom-right). In each panel, the cyan, blue, and orange lines show the results of EMBS, dynamical, AGN-MBH channels, respectively.}
\label{fig:ratio}
\end{figure}

The differences in the probability $P({\rm H}|{\rm GW})$ resulting from different sBBH formation channels suggest that these channels can be distinguishable via the detection rate of such events by future sky surveys. In the coming years, CSST, Euclid, and RST will survey a large fraction of the sky and find numerous lensed galaxies \citep[e.g.,][]{2013LRR....16....6A,2013arXiv1305.5422S,2019ApJ...883..203G}, which enable them to be able to identify the hosts of some lensed sBBH GW events. For example, if we adopt the i-band of CSST ($m_{\rm lim}=24.8$\,mag), VIS-band of Euclid ($m_{\rm lim}=24.5$\,mag), and J-band of RST ($m_{\rm lim}=27.0$\,mag), then the fraction of lensed sBBH mergers that CSST, Euclid, and RST can identify the lensing signatures of the hosts in its observational sky area can be roughly $0.11/0.08/0.25$, $0.09/0.07/0.21$, and $0.20/0.15/0.48$, respectively.  

The detection rate of the lensed sBBH events by the 3rd generation GW detectors (e.g., ET) has been estimated by a number of authors in the literature \citep[e.g.,][]{Piorkowska:2013eww, 2014JCAP...10..080B, 2018MNRAS.476.2220L, 2022MNRAS.509.3772Y}, in which the EMBS channel is mainly considered. This rate is predicted to be $\sim 11^{+12}_{-6}$\,yr$^{-1}$ for ET (re-scaled by the current local merger rate density estimate $\sim 16-61{\rm Gpc^{-3}\,yr^{-1}}$ with a mean of $29\,{\rm Gpc^{-3}\,yr^{-1}}$) according to \citet{2018MNRAS.476.2220L}, and it is roughly about $\sim 96^{+107}_{-43}$\,yr$^{-1}$ for the network of the 3rd generation GW detectors \citep[][also rescaled by the constrained local merger density]{2022MNRAS.509.3772Y}. The estimates for this detection rate may vary significantly if choosing a different sBBH formation channel because the merger rate density evolution and the chirp-mass distribution resulting from different channels are substantially different. For example, we estimate the detection rate of the lensed sBBH events by ET is $\sim 23-89$\,yr$^{-1}$ (or $649-2479$\,yr$^{-1}$) if assuming the dynamical (or AGN/MBH) channel as the dominant channel for the sBBH formation (also with the local merger rate density calibrated to $16-61{\rm Gpc^{-3}\,yr^{-1}}$), which are higher than that by assuming the EMBS channel. However, we note that there could be large uncertainties in the merger rate density evolution estimated in the present paper for the dynamical and AGN/MBH channels (see Fig.~\ref{fig:fig1}) and the estimated local merger rate densities could also be significantly different from the current constraint by LIGO/Virgo observations \citep[e.g.,][]{ 2020A&A...638A.119G,2021Symm...13.1678M}. Below we mainly consider the EMBS channel for the detection number estimation for different (survey) telescopes and it may be taken as a conservative estimate. For other channels, one may simply use the scaling to obtain the detection number. 

Some of the lensed sBBH merger hosts may be identified by the upcoming sky surveys, including CSST, Euclid, and RST. RST will survey $\sim 2000$\,deg$^2$ sky area within a five-year observation run and find $\sim 17,000$ strong gravitational lensed galaxies considering its survey strategy \citep[e.g.,][]{2020RNAAS...4..190W}. Regardless of the time when these lensed galaxies are found, one can always try to match the lensed GW events detected in the RST survey sky area with those lensed galaxies \citep[see][]{2020MNRAS.497..204Y,2022arXiv220408732W}. If adopting the RST survey in the J-band with the limiting magnitude of $27$\,mag, each year one would expect to find the lensed hosts of $\sim 0.11^{+0.12}_{-0.06}$ and $\sim 0.96^{+1.07}_{-0.43}$ lensed sBBH mergers detected by ET and the network of the 3rd generation GW detectors if assuming all lensed sBBHs are produced by the EMBS channel. CSST and Euclid will survey a sky area of $17,500$\,deg$^2$ and $15,000$\,deg$^2$, and will find $\sim 200,000$ and $170,000$ lensed galaxies, respectively. Considering the probability $P({\rm H}|{\rm GW})$ estimated for the i-band of CSST/the VIS-band of Euclid (see Fig.~\ref{fig:limit}), CSST and Euclid may be able to identify the lensing signatures of the hosts of $0.53^{+0.57}_{-0.29}$/$0.37^{+0.40}_{-0.20}$ lensed sBBH mergers per year by ET and $4.65^{+5.33}_{-2.08}$/ $ 3.26^{+3.59}_{-1.46}$ per year by the network of the 3rd generation GW detectors if assuming all lensed sBBHs are produced by the EMBS channel. 

\begin{figure}
\includegraphics[width=1.0\columnwidth]{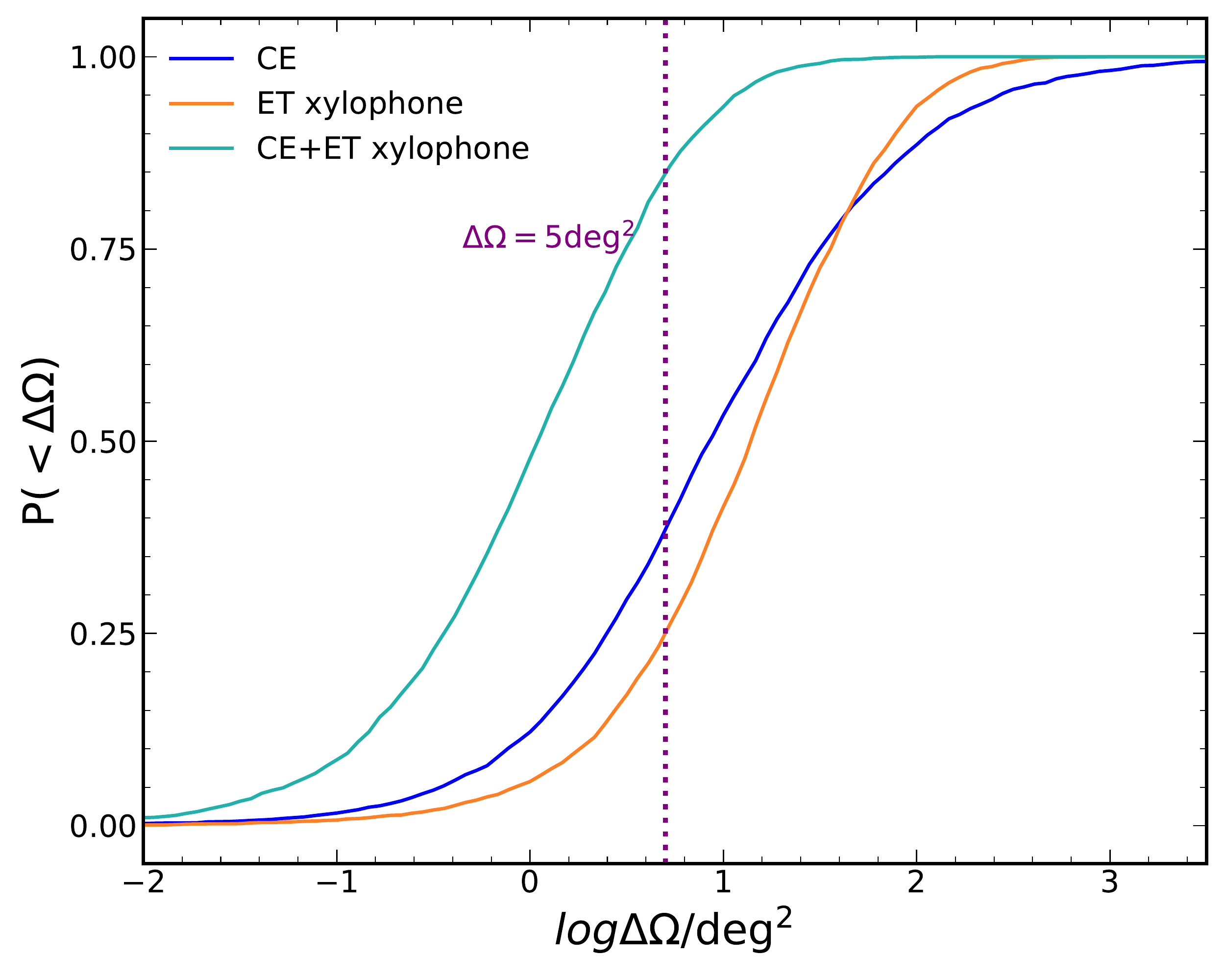}
\centering
\caption{The cumulative probability distribution of the GW localization errors for those lensed sBBH mergers detected by the 3rd generation ground-based GW detectors. The orange, blue, and cyan solid lines show the results obtained by the ET with xylophone design, CE, and the network of CE and ET with xylophone design, respectively. }
\label{fig:loc}
\end{figure}

Note that we assume that the exact host of a lensed GW event can always be identified once this host is among those galaxies with identifiable lensing signatures in the localization area of the GW event in the above analysis. Therefore, the above estimates of the rates can be only taken as the strict upper limits and one need to consider the localization errors of the detected lensed GW sBBH mergers to find the host galaxy. Figure~\ref{fig:loc} shows the cumulative probability distribution on the localization errors estimated from the Fisher matrix (see Eq.~\ref{eq:omega}) for those mock lensed GW events detected by different ground-based detectors, including ET with xylophone design, CE, and 3G network of detectors (CE+ET xylophone). As seen from this figure, ET-xylophone design or CE only can hardly localize those lensed sBBHs within a sky area less than a few square degree and thus it is difficult to identify the lensed hosts. With the powerful 3G GW network, however, $\sim 85\%$ of the mock lensed sBBH mergers can be localized within a sky area less than $5$ deg$^2$. If adding more GW detectors to the network as proposed by \citet{2018PhRvD..97f4031Z} and \citet{2019ApJ...887...28L}, the localization of the lensed sBBH mergers may be more precise. Therefore, as estimated by 
\citet{2022arXiv220408732W}, the above predicted rate may be reduced by at most a factor of $\sim 0.35-0.20$ due to this localization error of the GW signals ($1-5$ deg$^2$) and reconstruction errors for the lensed host galaxies.

Once both the lensed GW and host signals of a sBBH merger are detected, the exact location of the sBBHs can be localized by mapping the time-delay and magnification factors of the GW event to the lensed host \citep[e.g.,][]{2020MNRAS.498.3395H}. Thus one may obtain the statistical spatial distribution of the sBBH mergers in their hosts for those lensed sBBH mergers with identified hosts. Figure~\ref{fig:ratio} shows the spatial distribution of the lensed sBBH mergers in its host galaxies (scaled by the effective radius $R_{\rm e}$) identified by a survey at the i-band with  $m_{\rm lim}=24$\,mag (top-left), $25$\,mag (top-right), $26$\,mag (bottom-left), and $27$\,mag (bottom-right), respectively. As seen from this figure, assuming different sBBH formation channel result in different spatial distribution. It is obvious that those lensed sBBH mergers from the AGN-MBH channel are all located in the center of the hosts, while those from the dynamical and EMBS channels can exist at large radius, with the spatial distribution from the dynamical channel more extended than that from the EMBS channel. Adopting a Sersic-law [$\propto a+ b (R_{\rm GW}/R_{\rm e})^n$] to fit this spatial distribution at $< 5R_{\rm GW}/R_{\rm e}$, we find $b\sim -1.8$-$-2.0$/$-7.8$-$-11.0$ and $n\sim 0.3$/$1.0$ if assuming all the sBBH mergers are produced by the EMBS/dynamical channel. Some sBBH mergers in the cases assuming the dynamical channel may be located at $>5R_{\rm GW}/R_{\rm e}$, which is substantially distinct from that in the cases assuming the EMBS channel. However, these mergers may be hard to observe as the the fraction is very small. 

\section{Conclusions and discussions}
\label{sec:con}

In this paper, we show that the fraction of those lensed GW sBBH events that can have identifiable lensed host signatures mainly dependent on the origin of sBBHs and the limiting magnitude $m_{\rm lim}$ of the sky-surveys searching for it, which is $\sim 0.11/0.08/0.25$ for i-band of CSST, $\sim0.09/0.07/0.21$ for VIS-band of Euclid, and $\sim 0.20/0.15/0.48$ for J-band of RST if assuming that all the sBBH mergers are produced by the EMBS, dynamical, and AGN-MBH channels, respectively. In addition, we also demonstrate that the statistical distributions of those lensed sBBHs resulting from different sBBH formation channels can also be significantly different from each other. Therefore, one can distinguish different sBBH formation channels via the detection fraction of those lensed events with identifiable lensing host signatures and/or even constrain the contribution fractions from different formation channels. 

It has been shown that the gravitational lensed GW events can be used to accurately constrain the cosmological parameters via the precise time-delay measurements from the GW observations and the redshift measurements from EM observations \citep[e.g.][]{2017NatCo...8.1148L, PhysRevD.101.064011, 2020MNRAS.498.3395H}. For example, \citet{2017NatCo...8.1148L} showed that $H_0$ can be constrained to a precision $\sim 0.68\%$ if the EM counterparts of ten lensed GW events can be detected. \citet[][]{2020MNRAS.498.3395H} showed that $H_0$ can be constrained to a precision $\sim 10\%$ by only one event if assuming a $\sim 20\%$ scatter between the true magnifications of the GW events and its host galaxy (if the lensed host can be detected as discussed in this paper). In this paper, we find that at least about $0.4$ to $3$ lensed sBBH mergers per year, among those detected by the network of the 3rd generation GW detectors, can be identified with the lensed hosts revealed by sky surveys like CSST, Euclid, and RST, though challenge. After the accumulation of a decade, the number could be mounted up to ten or more and form a statistical robust sample and thus enable the application of gravitational lensed GW sources as a unique and independent probe to constrain the cosmological parameters. To get a larger sample one may also need sky surveys with the limiting magnitude much fainter than CSST and Euclid, which can more efficiently identify the hosts of the lensed GW events.

We note here that some simple assumptions and approximations made in this paper may affect the estimates quantitatively though do not change the general results. For demonstration purpose, we only consider simple cases in this paper. In reality, there are many complexities that one may need to take into account to make a more robust investigation. For example, we consider the simple SIE lens model with thin lens approximation when sampling the mock GW events and its hosts. We ignore several factors and put strict but rather flexible criteria on identifying both the lensed GW events and the lensed signals of the host galaxies for rough estimations, which may affect the actual detecting numbers for each survey.

We also made simple approximations for the intrinsic sBBH spatial distribution within host galaxies through concise physical considerations of different formation channels. As for the EMBS channel, we assume the mass-luminosity ratio as a constant, so that the luminosity distribution can represent the matter distribution in the host. However, it should evolve with redshift and depend on the radius via more comprehensive ways. For the dynamical channel, the spatial distribution of the sBBH mergers are assumed to follow the distributions of GCs in its hosts, which is simply approximated by the projected NFW profile. However, the distribution is much more complicated in a real  galaxy as it may depend on its assembly history and various properties (such as ages, metallicities, etc.).  Moreover, we treat these three channels separately by assuming sBBHs are produced via a single one, while all these channels may play a role at the same time for the sBBH formation.  
Note that we simply adopt the Fisher information matrix method as a rough estimation on the localization precision of lensed sBBH mergers, which requires
for the satisfaction of the unbiased and high SNR approximation \citep[e.g.,][]{2008PhRvD..77d2001V}. This condition may not be always satisfied in real detections of lensed GW signals. Therefore, to make a more robust investigation, it is encouraging to conduct parameter estimation within the framework of Bayesian approach \citep[e.g.][]{2009PhRvD..79l4033R,2012PhRvD..85j4045V, 2014ApJ...789L...5K, 2014ApJ...795..105S, 2014PhRvD..89d2004G, 2015PhRvD..91d2003V, 2015ApJ...804..114B}, for example, with the use of several reliable analyzing tools, such as \textbf{\textsc{LALinference}} \citep[e.g., ][]{2015PhRvD..91d2003V, 2015ApJ...804..114B}. Moreover, most lensed sBBH mergers can only be constrained within $\sim \rm 5 deg^{2}$ as shown in Section~\ref{sec:results}, within which upto several tens lens systems can be observed by future sky surveys  \citep[e.g., ][]{Collett_2015,2020RNAAS...4..190W}. To find the real host, it is necessary to match those candidate lensed hosts with the lensed sBBH mergers via the mapping of the time delay(s) and magnification factor ratio(s), which is strongly dependent on the reconstruction of the host galaxy image. Therefore, there are many complexities one may need to further take into account to more accurately estimate the detection rate of lensed sBBH mergers with identifiable lensed host galaxies.

Note also that we limit our analysis in the present paper to the galaxy-galaxy lensing but ignore the galaxy-cluster lensing simply because the cluster lensing is rarer \citep[e.g.,][the relative rate of lensed events by cluster is at most half of that estimated by galaxy-galaxy lensing]{2018MNRAS.475.3823S, 2022arXiv220412977S, 2021ApJ...923...14A}. Furthermore, the time delay and flux ratio distributions for galaxy cluster lenses are more difficult to model accurately comparing with those for the galaxy lenses, owing to the more complex lensing morphology. Thus it may be hard to localize the positions of sBBH mergers in its hosts for cluster lensing systems.

\section*{acknowledgement}
This work is partly supported by the National Natural Science Foundation of China (Grant No. 12273050, 11690024, 11873056, 11991052), the Strategic Priority Program of the Chinese Academy of Sciences (Grant No. XDB 23040100), and the National Key Program for Science and Technology Research and Development (Grant No. 2020YFC2201400 and 2016YFA0400704).

\bibliographystyle{aasjournal}
\bibliography{ref.bib}

\end{document}